\documentclass[conference]{IEEEtran}
\usepackage[utf8]{inputenc}
\usepackage{graphicx}
\usepackage{color}
\usepackage{balance}

\begin{document}

\title{Towards Utility-based Prioritization of Requirements in Open Source Environments}

\author{\IEEEauthorblockN{Alexander Felfernig\IEEEauthorrefmark{1}, Martin Stettinger\IEEEauthorrefmark{1}, Müslüm Atas\IEEEauthorrefmark{1}, Ralph Samer\IEEEauthorrefmark{1},\\
Jennifer Nerlich\IEEEauthorrefmark{2}, Simon Scholz\IEEEauthorrefmark{2}, Juha Tiihonen\IEEEauthorrefmark{3}, and Mikko Raatikainen\IEEEauthorrefmark{3}\\}
\IEEEauthorblockA{\IEEEauthorrefmark{1}Graz University of Technology, Austria, Email: \{afelfern,mstettinger,muatas,rsamer\}@ist.tugraz.at,\\}
\IEEEauthorrefmark{2}vogella GmbH, Hamburg, Germany, Email: \{jennifer.nerlich,simon.scholz\}@vogella.com,\\
\IEEEauthorrefmark{3}University of Helsinki, Finland, Email: \{juha.tiihonen,mikko.raatikainen\}@helsinki.fi
}
\maketitle

\maketitle

\begin{abstract}
Requirements Engineering in open source projects such as \textsc{Eclipse} faces the challenge of having to prioritize requirements for individual contributors in a more or less unobtrusive fashion. In contrast to conventional industrial software development projects, contributors in open source platforms can decide on their own which requirements to implement next. In this context, the main role of prioritization is to support contributors in figuring out the most relevant and interesting requirements to be implemented next and thus avoid time-consuming and inefficient search processes. In this paper, we show how utility-based prioritization approaches can be used to support contributors in conventional as well as in open source Requirements Engineering scenarios. As an example of an open source environment, we use  \textsc{Bugzilla}. In this context, we  also show how dependencies can be taken into account in utility-based prioritization processes.
\end{abstract}


\IEEEpeerreviewmaketitle

\section{Introduction}
In software projects, resources are typically limited which requires the prioritization of requirements \cite{Lehtola2004}. Prioritization is often interpreted as a part of \emph{strategic planning} where the  focus is to select and prioritize requirements that should be included in   releases (long-term release planning) \cite{Ameller2017,88_Ruhe2005}. Decision support in prioritization is extremely important since especially when dealing with large assortments of requirements, manual prioritization processes tend to become very costly \cite{Alenezi2013,Xuan2012}. In this context, suboptimal prioritizations trigger time wasting due to the implementation of unimportant requirements.

There are \emph{two basic approaches} to prioritize requirements -- for an in-depth related analysis we refer to Achimugu et al. \cite{Achimugu2014}. \emph{First}, requirements prioritization can be interpreted as an \emph{optimization task} where the overall objective is to identify the \emph{middle ground}, i.e., an aggregation of individual prioritizations into a global prioritization that reflects the least possible level of dissimilarity from all stakeholder-individual prioritizations \cite{Kifetew2017}. \emph{Second}, in contrast to approximating individual prioritizations on the basis of optimization functions, \emph{utility-based approaches}  focus on (1) establishing agreement with regard to the evaluation of individual requirements and (2) thereafter determining prioritizations \cite{Adomavicius2010,Huang2011,Wiegers2003}. 

Prioritizations following the \emph{optimization approach} are determined on the basis of individual prioritizations of stakeholders. When following a \emph{utility-based approach}, preferences of stakeholders are first aggregated and a prioritization is determined thereafter. In the line of basic approaches to determine group recommendations \cite{Felfernig2018}, the first approach is based on \emph{aggregated prioritizations}  where stakeholder-individual prioritizations are known and a recommendation minimizes dissimilarities between the given prioritizations (preferences). The second approach is based on \emph{aggregated models} where stakeholder requirement evaluations are aggregated first and a prioritization is determined on the basis of a group profile (model) derived from requirements evaluations. 

Aggregated models have the advantage that stakeholders are encouraged to focus their evaluations on specific relevant aspects of a requirement (e.g., dimensions such as profit, risk, and effort) and thus contribute to stable preferences and a higher degree of consensus \cite{Stettinger2015IUI}. Aggregated prioritizations trigger scalability issues since each stakeholder has to provide, for example, a ranked list of requirements as input for the optimization process. Furthermore, due to the computational complexity of the underlying problem, an optimal solution cannot be guaranteed and is often only approximated on the basis of local search algorithms \cite{Kifetew2017,Tonella2013,Zhang2007}. Utility-based approaches as discussed in this paper focus on evaluations of individual requirements on the basis of different evaluation dimensions (e.g., profit, effort, and risk). This way, stakeholders can focus on evaluating requirements they have knowledge about and the focus of prioritization is first on establishing \emph{consensus} and thereafter on figuring out the most relevant prioritizations \cite{Stettinger2015IUI}.

Different algorithmic approaches can be used to support requirements prioritization -- for an overview, see, for example, Achimugu et al. \cite{Achimugu2014}. Examples thereof are constraint-based reasoning \cite{Tsang1993}, incremental preference learning \cite{Perini2013}, evolutionary algorithms \cite{Kifetew2017}, machine learning \cite{Alenezi2013,Tian2015}, and pairwise preference-based decision making \cite{Saaty2000}. \emph{Optimization-based approaches} focus on minimizing the distance between the preferences of individual stakeholders (e.g., in terms of the distance between individual prioritizations and the prioritization determined by the optimization approach). A similar problem also solved on the basis of optimization approaches is the \emph{next release problem} \cite{Bagnall2001,Xuan2012nextrelease} where a subset of a given set of requirements has to be selected in such a way that predefined cost limits are taken into account and the chosen set of requirements represents the optimum choice in terms of criteria such as market value. The focus in this context is more to identify subsets of requirements but not to prioritize a given list of requirements. In contrast to next release problems, prioritization tasks in open source scenarios do not necessarily require (and often do not allow) a global optimum but more focus on relevant recommendations for individual stakeholders. Also in contrast to existing release planning tasks, developers in open source scenarios in most of the cases do not explicitly define their preferences, i.e., preferences have to derived from given interaction data (in our case, interaction data collected by \textsc{Bugzilla}\footnote{www.bugzilla.org.}).

\emph{Utility-based prioritization} based on  \emph{multi-attribute utility theory} \cite{Dyer2005} can be implemented in different variants. First, requirements are simply evaluated with regard to a set of predefined interest dimensions and the overall utility of a requirement is determined as a sum of  interest dimension specific utilities. Second, weights can be introduced to emphasize on specific interest dimensions (e.g., a lower risk is more important than high profits). Third, stakeholders can be enabled to define their personal evaluations and utility-based approaches should then be able to aggregate these evaluations and take into account stakeholder weights. Stakeholder weights can be interpreted as "global", i.e., there is a global weighting of stakeholders independent of a specific dimension or requirement. If weights are interpreted as "local", the importance of a stakeholder can be defined on the level of individual requirements or dimensions. Utility-based prioritization can also be implemented on the basis of analytic hierarchy process (AHP) \cite{Karlsson1997}. A major disadvantage of this approach is that requirements have to evaluated pairwise which does not scale well when the number of requirements increases.

Prioritization criteria differ depending on the requirements engineering scenario. The criteria \emph{effort}, \emph{risk}, and \emph{profit} are often used in settings where a group of stakeholders  engaged in the same project is in charge of completing a prioritization task \cite{Achimugu2014}. In contrast, in open-source settings, developers are in most of the cases engaged in different projects and also work for different companies. In such scenarios, prioritization is less focusing on establishing consensus between individual stakeholders but more on supporting stakeholders in identifying requirements of relevance to them and to prioritize the important ones by also taking into account global criteria. Examples of criteria in such scenarios are \emph{personal expertise} of a developer and \emph{importance} of a requirement for the community of the stakeholder and the open source community as a whole. Thus, open source platform related prioritization processes completely differ from conventional software projects. A major focus of this paper is to introduce prioritization concepts especially applicable in open source development contexts.

\textsc{Bugzilla} is an open-source based issue tracking system which supports users  from different geographical locations to report their findings with regard to a given set of software components. Users can submit textual descriptions of issues and corresponding meta-information, for example, associated components, keywords, and dependencies. Reported issues can be selected by contributors to work on. In \textsc{Bugzilla}, issues can be requirements but also reported bugs. Distinguishing between these can be performed on the basis of a meta-attribute (issue type) that can be specified for \textsc{Bugzilla} issues. There are different related approaches to support machine learning based requirements prioritization. The approaches operate on datasets including historical data of previous requirement (bug report in \textsc{Bugzilla}) selections and try to predict future requirement selections thereof. Utility-based prioritization can be used in interactive scenarios (stakeholders are engaged in an interactive prioritization process) as well as scenarios where requirements are recommended but no further stakeholder interaction is needed for determining a prioritization.

The \emph{contributions} of this paper are the following. We provide an overview of different application scenarios of utility-based requirements prioritization and discuss specific aspects of requirements prioritization in open source projects. For scenarios that include dependencies between requirements, we show how such dependencies can be taken into account on the basis of the concepts of model-based diagnosis \cite{FelfernigSchubertZehentner2012,84_Reiter1987}. With this approach, we tackle the following \emph{research gaps}. In contrast to existing prioritization and release planning approaches, we introduce model-based diagnosis concepts that also support re-prioritization and re-planning while not completely omitting already existing stakeholder preferences which is still an open issue in most of the existing prioritization and release planning approaches (these approaches focus on taking into account dependencies but do not support the aforementioned re-prioritization and re-planning scenarios). Furthermore, we present a first version of a user interface developed to support prioritization tasks in open source environments -- in our case, for \textsc{Bugzilla} users. This approach has the potential to reduce the workload of developers in open source platforms by automatically proposing requirements that should be implemented next instead of forcing users to analyze in detail a large number of requirements. Furthermore, the introduced utility-based approach does not require the "manual" evaluation of individual requirements but automatically derives utility models by analyzing given interaction logs taking into account interaction data such as \emph{number of comments} related to a requirement. This is a major difference compared to existing requirements prioritization approaches which do not support automated prioritization based on background data. Finally, we discuss issues for future work to further advance the state of the art in  utility-based prioritization.

The remainder of this paper is organized as follows. In Section \ref{utilitybasedprioritization}, we introduce variants of implementing \emph{utility-based prioritization}. Thereafter, we introduce our variant of utility-based prioritization implemented for the  \textsc{Bugzilla} environment and provide a sketch of a related \textsc{Bugzilla} user interface (Section \ref{bugzillaprioritization}).  In Section \ref{dependencies}, we show how to extend utility-based prioritization in such a way that dependencies between requirements can be taken into account. The paper is concluded with a discussion of future work in Section \ref{conclusion}.

\section{Utility-based Prioritization}\label{utilitybasedprioritization}
Utility-based prioritization allows stakeholders to prioritize a requirement with regard to different interest dimensions $D = \{d_1, d_2, ..., d_n\}$. Examples of such interest dimensions are \emph{profit}, \emph{risk}, and \emph{effort}. Utility-based prioritization is based on the idea to first evaluate each requirement with regard to the set of interest dimensions (see Table \ref{basicevaluationsetting}) and thereafter calculate the individual utility of each requirement (see Formula \ref{staticpriority}). In general, the \emph{priority} is associated with the \emph{utility} of a requirement $r$ which results from its total contributions to all of each individual interest dimensions $d$ (denoted as $contribution(r,d)$) combined with the corresponding importance weights of individual interest dimensions (denoted as $weight(d)$).

\begin{table}[!ht]
\caption{Contribution of requirements $R=\{r_1,r_2,r_3\}$ to the interest dimensions $D=\{profit, risk, effort\}$.} \label{basicevaluationsetting}
\centering{}\begin{tabular}{|c|c|c|c|c|c|} 
\hline 
interest dimension      	& $r_1$ 	& $r_2$  	& $r_3$        \tabularnewline
\hline
\hline
profit 				& 10 		& 5 		& 4       \tabularnewline
\hline
risk 				& 7 		& 2 		&  8       \tabularnewline
\hline
effort 				& 2 		& 3 		& 7       \tabularnewline
\hline
\end{tabular}  
\end{table}

\begin{table}[!ht]
\caption{Predefined weights for the interest dimensions $D=\{profit, risk, effort\}$.}
\centering{}\begin{tabular}{|c|c|c|c|c|c|} 
\hline 
interest dimension      	& weights  \tabularnewline
\hline
\hline
profit 				& 0.3 		      \tabularnewline
\hline
risk 				& 0.5 		      \tabularnewline
\hline
effort 				& 0.2 		      \tabularnewline
\hline
\end{tabular}  
 \label{singleuserpreferences}
\end{table}

\begin{equation}\label{staticpriority}
utility(r,D) = \Sigma_{d \in D} (contribution(r,d) \times weight(d))
\end{equation}

\vspace{0.5cm}

Applying Formula \ref{staticpriority} to the entries in Tables \ref{basicevaluationsetting} and \ref{singleuserpreferences} results in the ranking depicted in Table \ref{basicevaluationranking} (the higher the utility with regard to the given interest dimensions, the higher the corresponding priority of the requirement).

\vspace{0.5cm}

\begin{table}[!ht]
\caption{Ranking of requirements with static weights.}
\centering{}\begin{tabular}{|c|c|c|c|c|c|} 
\hline 
requirement      	& $r_1$ 	& $r_2$  	& $r_3$        \tabularnewline
\hline
\hline
utility 			& 6.9 		& 3.1 		& 6.6       	\tabularnewline
\hline
priority (ranking) 	& 1 		& 3 		& 2       	\tabularnewline
\hline
\end{tabular}  
 \label{basicevaluationranking}
\end{table}

\begin{table*} [!ht]
\caption{Contribution of requirements $R=\{r_1,r_2,r_3\}$ to dimensions $D=\{profit, risk, effort\}$ (defined by stakeholders $S = \{s_1,s_2,s_3\}$).}
\center
\begin{tabular}{|l|c|c|c|c|c|c|c|c|c|c|c|c|c|c|c|c|} \hline
interest    &\multicolumn{4}{  c  |}{$r_1$}  &\multicolumn{4}{  c  |}{$r_2$} &\multicolumn{4}{  c  |}{$r_3$} \\ \cline{2-13}   
dimension   & $s_1$  & $s_2$ & $s_3$ & $AVG$		& $s_1$  & $s_2$ & $s_3$ & $AVG$	 & $s_1$  & $s_2$ & $s_3$ & $AVG$ \\ \hline  \hline
 profit 	&	5    & 	2 	 & 	2 & 3.0   			&	5    & 	1 	 & 	2 & 2.7  &	2    & 	2 	 & 	6 & 3.3		\\ \hline   
 risk 		&	3    & 	3 	 & 	4 & 3.3   			&	2    & 	5 	 & 	6 & 4.3	 &	3    & 	2 	 & 	2 & 2.3		\\ \hline  
 effort 	&	2    & 	3 	 & 	2 & 2.3   			&	3    & 	4 	 & 	2 & 3.0	 &	5    & 	6 	 & 	2 & 4.3		\\ \hline  
\end{tabular}

\label{groupevaluation}
\end{table*}

In the previous example, the evaluation of requirements with regard to interest dimensions and the weighting of interest dimensions are assumed to be predefined (e.g., by a single stakeholder). However, requirements prioritization is often a group decision process \cite{Felfernig2018} where different stakeholders are evaluating requirements (see, e.g., Table \ref{groupevaluation}) and define importance weights with regard to interest dimensions (see, e.g., Table \ref{groupweights}). Both, stakeholder-individual evaluations of interest dimensions and importance weights have to be aggregated. Formula \ref{groupcontribution} shows the aggregation of stakeholder-individual evaluations of requirements where $S$ refers to the set which includes all $m$ stakeholders (i.e., $S = \{s_1, s_2, ..., s_m\}$).

\vspace{0.25cm}

\begin{equation}\label{groupcontribution}
contribution(r,d,S) = \frac{\Sigma_{s \in S} eval(d,r,s)}{|S|}
\end{equation}

\vspace{0.5cm}

Formula \ref{groupweightweightedstakeholders} shows how to aggregate the stakeholder-specific importance weights (denoted as $w(d,s)$) which are related to individual interest dimensions $d$. Previous calculations did not take into account potential different degrees of stakeholder expertise, for example, a stakeholder $s_a$ could have more expertise with regard to estimating the market potential of a requirement in terms of profit as estimating the corresponding development efforts. To take into account this aspect, Formula \ref{groupweightweightedstakeholders} includes a factor that represents the \emph{expertise} of a stakeholder $s$ with regard to a specific interest dimension $d$.

\begin{equation}\label{groupweightweightedstakeholders}
weight(d,S) = \frac{\Sigma_{s \in S} w(d,s) \times expertise(d,s)}{|S|}
\end{equation}

\vspace{0.25cm}

Similar to the basic approach, the utility of a requirement (Formula \ref{prioritygroup}) is determined as a combination of the contributions of a requirement to the given interest dimensions and related interest dimension importance evaluations of stakeholders.

\begin{equation}\label{prioritygroup}
utility(r,D,S) = \Sigma_{d \in D} (contribution(r,d, S) \times weight(d,S))
\end{equation}


\begin{table}[!ht]
\caption{Preferences of stakeholders $S = \{s_1, s_2, s_3\}$ with regard to the interest dimensions $D=\{profit, risk, effort\}$.}
\centering{}\begin{tabular}{|c|c|c|c|c|c|} 
\hline 
stakeholder      	& $s_1$ 	& $s_2$ 	& $s_3$		& weights      \tabularnewline
\hline
\hline
profit 				& 0.5 		& 0.3		& 0.6	    & 0.47  \tabularnewline
\hline
risk 				& 0.3 		& 0.6 		& 0.3	    & 0.4 \tabularnewline
\hline
effort 				& 0.2 		& 0.1 		& 0.1	    & 0.13 \tabularnewline
\hline
\end{tabular}  
 \label{groupweights}
\end{table}

The result of applying Formulae \ref{groupcontribution}--\ref{prioritygroup} to the evaluation data contained in Tables \ref{groupevaluation} and \ref{groupweights} is depicted in Table \ref{groupevaluationranking}.

\begin{table}[!ht]
\caption{Ranking of requirements with group weights.}
\centering{}\begin{tabular}{|c|c|c|c|c|c|} 
\hline 
requirement      	& $r_1$ 	& $r_2$  	& $r_3$        \tabularnewline
\hline
\hline
utility 			& 3.03 		& 3.57 		& 3.03       	\tabularnewline
\hline
priority (ranking)	& 2 		& 1 		& 2       	\tabularnewline
\hline
\end{tabular}  
 \label{groupevaluationranking}
\end{table}

\section{Utility-based Prioritization in \textsc{Bugzilla}}\label{bugzillaprioritization}
In Section \ref{utilitybasedprioritization}, we took a look at different variants of utility-based prioritization. These variants were discussed on the basis of interest dimensions (evaluation criteria) typically occurring in software projects where a group of stakeholders is in charge of jointly defining and prioritizing requirements. In this section, we focus on open source scenarios where individual users (e.g., contributors in an open source platform) follow their individual interests regarding requirements without necessarily taking into account the preferences of other users. This can be considered a major difference compared to conventional software projects where stakeholders commonly develop a "global" prioritization (see Section \ref{utilitybasedprioritization}). We now show how utility-based prioritization can be applied in such contexts.

Table \ref{groupevaluationbugzilla} represents a \textsc{Bugzilla}-specific evaluation of requirements (bugs) with regard to the set of interest dimensions $\{cc, geritt, blocker, comments\}$. In this context, $cc$ is the number of contributors who are in the :cc list of bug-related emails, $geritt$ is the number of bug-related \textsc{Geritt}\footnote{A code reviewing tool -- gerritcodereview.com.} changes, $blocker$ is the number of dependent bugs (dependent requirements), and $comments$ refers to the number of bug-related comments. These interest dimensions do need to be evaluated manually as it is often the case in other scenarios \cite{Laurent2007,Shao2017} but can directly be determined from corresponding user interaction data.

Formula \ref{groupcontributionbugzilla} supports the calculation of the contribution of a requirement $r$ to a specific interest dimension $d$. In sharp contrast to the previous scenarios, the contribution is not directly specified by stakeholders but derived from \textsc{Bugzilla} specific information (e.g., \#comments related to a requirement).

\begin{equation}\label{groupcontributionbugzilla}
contribution(r,d) = eval(r,d)
\end{equation}

\vspace{0.25cm}

Formula \ref{groupweightweightedstakeholdersbugzilla} supports the determination of the expertise of a stakeholder which is represented in terms of the similarity between the keywords stored in the stakeholder profile and those extracted from the requirement description and corresponding meta-information (e.g., name of associated component/system). The similarity between requirement-related keywords, meta-information, and  contributor profile information is interpreted as \emph{expertise} (see Formula \ref{groupweightweightedstakeholdersbugzilla}). 

\begin{equation}\label{groupweightweightedstakeholdersbugzilla}
weight(r,s) = expertise(r,s)
\end{equation}

\vspace{0.25cm}

In the line of the previously discussed utility functions, the overall utility of a requirement is interpreted as a combination of (1) the contributions of a requirement to a set of interest dimensions and (2) the expertise level of a stakeholder (in this context interpreted "globally", i.e., not on the level of individual interest dimensions). 

\begin{equation}\label{prioritygroupbugzilla}
utility(r,s) = \Sigma_{d \in D} contribution(r,d) \times weight(r,s)
\end{equation}

\begin{table} [!ht]
\caption{Contribution of requirements (bugs) $R=\{r_1,r_2,r_3\}$ to the interest dimensions $D=\{cc, geritt, blocker,comments\}$.}
\center
\begin{tabular}{|l|c|c|c|c|c|c|c|c|c|c|c|c|c|c|c|c|} \hline
interest dimension   &\multicolumn{1}{  c  |}{$r_1$}  &\multicolumn{1}{  c  |}{$r_2$} &\multicolumn{1}{  c  |}{$r_3$} \\ \hline \hline
 cc 		&	5    & 	2 	 & 	2 	\\ \hline   
 geritt 	&	3    & 	3 	 & 	4 	\\ \hline  
 blocker 	&	2    & 	3 	 & 	2 	\\ \hline  
 comments 	&	2    & 	3 	 & 	2 	\\ \hline  
\end{tabular}

\label{groupevaluationbugzilla}
\end{table}

\begin{table}[!h]
\caption{Expertise of stakeholder $s_1$ with regard to the requirements $\{r_1, r_2, r_3\}$ determined, for example, on the basis of the similarity between the stakeholder profile and information associated with a requirement.}
\centering{}\begin{tabular}{|c|c|c|c|c|c|} 
\hline 
stakeholder      	& $s_1$ 		\tabularnewline
\hline
\hline
$r_1$ 				& 0.5 			\tabularnewline
\hline
$r_2$ 				& 0.3 			\tabularnewline
\hline
$r_3$ 				& 0.2 			\tabularnewline
\hline
\end{tabular}  
 \label{groupweightsbugzilla}
\end{table}

\begin{table}[!h]
\caption{Ranking of \textsc{Bugzilla} bugs with static weights.}
\centering{}\begin{tabular}{|c|c|c|c|c|c|} 
\hline 
requirement      	& $r_1$ 	& $r_2$  	& $r_3$        \tabularnewline
\hline
\hline
utility 			& 6.0 		& 3.3 		& 2.0       	\tabularnewline
\hline
priority 			& 1 		& 2 		& 3       		\tabularnewline
\hline
\end{tabular}  
 \label{staticrankingbugzillabugs}
\end{table}

\begin{figure*} [ht!]
\caption{\textsc{Bugzilla} view on bugs (requirements). Based on the presented utility-based prioritization approach, bugs are presented to \textsc{Bugzilla} contributors in the order of the determined priority.}
\begin{centering}
\includegraphics[scale=0.725]{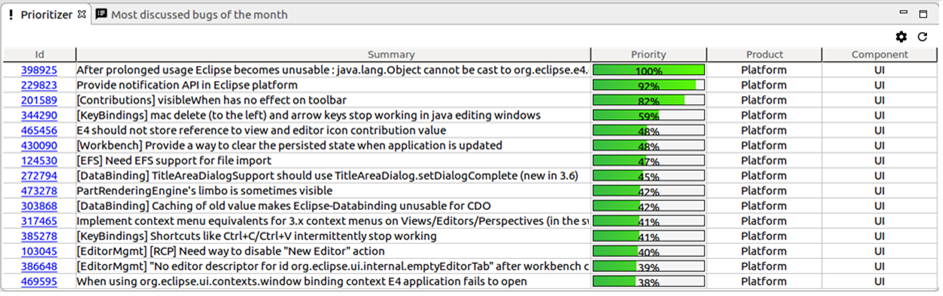}
\par\end{centering}
 \vspace{-0.45cm}
\label{fig:smarthomesimplified} 
\end{figure*}

\section{Taking into Account Dependencies}\label{dependencies}

\emph{Blocking Factor}. Utility-based recommendation approaches per se do not explicitly take into account dependencies between requirements (e.g., requirement $A$ must be implemented before requirement $B$). In the discussed open source prioritization scenario, this aspect is taken into account by prioritizing requirements on the basis of the number of related dependencies, i.e., the higher the number of requirements dependent from a requirement $x$, the higher the "global" relevance of $x$. In this context, the \emph{blocking factor} (i.e., how many requirements depend on the implementation of requirement $x$) can be considered as interest dimension that has an impact on prioritization. In other words, this requirement should be implemented as soon as possible since it otherwise blocks the implementation of other requirements. This approach can also be applied in software development scenarios where a  group of stakeholders (e.g., an in-house software development project) is in charge of prioritizing requirements. Such an approach helps to avoid situations where prioritizations violate dependency constraints.  

\emph{Automated Repair}. An alternative approach is to apply repair mechanisms from model-based diagnosis \cite{FelfernigSchubertZehentner2012} that help to adapt already determined prioritizations in such a way that all defined dependencies are taken into account. In the following, we will shortly sketch our approach. In order to trigger a diagnosis process, we are in the need of a pre-defined set of dependencies between requirements (denoted as $DEP = \{dep_1, dep_2, .., dep_n\}$). Furthermore, we assume that a prioritization (represented as sequence) $P = [p_1, p_2, .., p_m]$ determined by a utility-based prioritization approach is \emph{inconsistent} with the given set of dependencies. In order to apply model-based diagnosis, we assume that both, the pre-defined set of dependencies and the requirement prioritization is represented in terms of constraints \cite{Tsang1993}, for example, $DEP = \{dep_1: r_3 < r_1, dep_2: r_3 < r_2\}$ and $P = \{p_1: r_1 < r_2, p_2: r_2 < r_3, p_3: r_3 < r_4, p_4: r_4 < r_5, p_5: r_5 < r_6\}$. As can be easily seen, $DEP \cup P$ is inconsistent. As variable domains we assume $[1..\#requirements]$.

Following the principles of model-based diagnosis \cite{FelfernigSchubertZehentner2012,84_Reiter1987}, we need to detect all \emph{minimal conflicts} \cite{52_Junker2004} induced in $P$ by the dependencies defined in $DEP$. In this context, a conflict set is defined as follows.

\emph{Definition: Conflict Set (CS)}. A conflict set $CS \subseteq P$ is a set of individual prioritization elements that are inconsistent with the elements of $DEP$, i.e., \emph{inconsistent}($CS \cup DEP$). A conflict set $CS$ is minimal if $\neg \exists CS': CS' \subset CS$.

On the basis of a set of identified minimal conflict sets, a corresponding diagnosis includes a set of prioritization elements in $P$ that have to be adapted such that the consistency of $DEP \cup P$ is restored (see the following definition).

\emph{Definition: Diagnosis ($\Delta$)}. A diagnosis $\Delta \subseteq P$ is a set of individual prioritization elements that have to be deleted from $P$ such that \emph{consistent}($P - \Delta \cup DEP$).

In our example, $CS: \{p_2\}$ is the only conflict induced in $P$ by the dependencies defined in $DEP$. $CS$ is minimal, i.e., we need to adapt only one of the prioritization elements in $CS$ such that a global prioritization can be found that is consistent with the elements in $DEP$ \cite{52_Junker2004}. A corresponding diagnosis $\Delta$ is $\{p_2\}$. In our example, we could decide to replace $p_1: r_1 < r_2$ with the corresponding repair $r_1 > r_2$. This is a repair action that helps to restore the consistency of $DEP \cup P$.

Our approach to the repair of inconsistent prioritizations can be used for both, \emph{interactive prioritization} where stakeholders receive feedback on the consistency of prioritizations, and \emph{automated prioritization} where repairs for inconsistent prioritizations are determined in an automated fashion. Important issues to improve our approach are discussed in the following.

\section{Conclusion and Future Work}\label{conclusion}

\emph{Conclusion}. In this paper, we showed how to support utility-based requirements prioritization. These scenarios range from \emph{single user prioritization} where one stakeholder is in charge of completing prioritization tasks to \emph{group-based prioritization} where the preferences/evaluations of different group members have to be taken into account. On the basis of these scenarios, we showed how utility-based prioritization can be applied in the context of open source development projects. In this context, we sketched our initial implementation currently provided in the \textsc{Bugzilla} environment. This implementation serves as a first version to support prioritization in \textsc{Bugzilla}. 

\emph{Future Work}. Since prioritization is a repetitive process, we will include mechanisms that are capable of learning stakeholder weights and also the weights of individual requirements. This approach will help to further increase the prediction quality of prioritizations in terms of the probability that stakeholders accept the proposed prioritizations. In this context, we will also compare the predictive quality of utility-based approaches (i.e., approaches based on aggregated models) with machine learning based approaches and approaches that determine rankings on the basis of aggregated prioritizations. Furthermore, we will analyze which further features (interest dimensions) are useful to improve prediction quality. For example, the number of redundant bugs (issues) can be a further important relevance indicator. 

A major challenge in requirements prioritization is the provision of persuasive user interfaces that increase the preparedness of stakeholders to actively engage in requirements engineering processes \cite{Atas2017}. Consequently we will focus on a further extension/improvement of the existing \textsc{Bugzilla} requirements prioritization user interface. Furthermore, we will analyze in which way recommended prioritizations have to be \emph{explained} to support specific group decision goals such as \emph{consensus}, \emph{fairness}, and \emph{decision quality} \cite{Du2009,Felfernig2018}. Finally, we will conduct a detailed empirical study regarding the impact of our prioritization approaches in the \textsc{Eclipse} community.

\balance

\section*{Acknowledgment}
The work presented in this paper has been conducted within the scope of the \textsc{WeWant} Project (funded by the Austrian Research Promotion Agency) and the Horizon 2020 Project \textsc{OpenReq}.

\bibliographystyle{plain}
\bibliography{sigproc}

\balance

\end{document}